# Dense polarized positrons from beam-solid interactions

**Authors:**

Xing-Long Zhu[1,2*], Wei-Yuan Liu[3,4], Tong-Pu Yu[5], Min Chen[3,4], Su-Ming Weng[3,4], Wei-Min Wang[6,4*], and Zheng-Ming Sheng[2,3,4*]

**Affiliations:**

[1] Institute for Fusion Theory and Simulation, School of Physics, Zhejiang University, Hangzhou 310058, China

[2] Tsung-Dao Lee Institute, Shanghai Jiao Tong University, Shanghai 200240, China

[3] Key Laboratory for Laser Plasmas (MOE), School of Physics and Astronomy, Shanghai Jiao Tong University, Shanghai 200240, China

[4] Collaborative Innovation Center of IFSA, Shanghai Jiao Tong University, Shanghai 200240, China

[5] Department of Physics, National University of Defense Technology, Changsha 410073, China

[6] Department of Physics and Beijing Key Laboratory of Opto-electronic Functional Materials and Micro-nano Devices, Renmin University of China, Beijing 100872, China

[*]Email: xinglong.zhu@zju.edu.cn (X.L.Z.); weiminwang1@ruc.edu.cn (W.M.W.); and zmsheng@sjtu.edu.cn (Z.M.S.)

**Abstract:** Relativistic positron sources with high spin polarization have important applications in nuclear and particle physics and many frontier fields. However, it is challenging to produce dense polarized positrons. Here we present a simple and effective method to achieve such a positron source by directly impinging a relativistic high-density electron beam on the surface of a solid target. During the interaction, a strong return current of plasma electrons is induced and subsequently asymmetric quasistatic magnetic fields as high as megatesla are generated along the target surface. This gives rise to strong radiative spin flips and multiphoton processes, thus leading to efficient generation of copious polarized positrons. With three-dimensional particle-in-cell simulations, we demonstrate the production of a dense highly-polarized multi-GeV positron beam with an average spin polarization above 40% and nC-scale charge per shot. This offers a novel route for the studies of laserless strong-field quantum electrodynamics physics and for the development of high-energy polarized positron sources.

The creation of relativistic positrons is a critical phenomenon in strong-field quantum electrodynamics (QED) physics and is fundamentally interesting for broad research areas [1-5]. In particular, when the positrons carry spin polarization, they could be applied to a few fundamental problems related to our universe, such as searching new physics beyond the standard model [6], probing nucleon structures [7], and understanding extreme astrophysical phenomena related to black holes [8], $\gamma$-ray bursts [9] and pulsar magnetospheres [10]. Although high-density, high-energy polarized positrons are believed to be ubiquitous in many energetic astrophysical objects [8-10], they are still difficult to obtain in laboratory.

Considerable efforts have been made to attain polarized positrons with accelerators [11-13]. For example, relativistic positrons in storage rings can be polarized via Sokolov-Ternov effect [14, 15], which, however, has strict requirements on experimental operation and often is time consuming. Currently, there are two commonly used alternative methods to generate polarized electron beams based upon the Bethe-Heitler process, which are achieved by use of circularly polarized $\gamma$-photon beams [16, 17] or polarized electron beams [18] interacting with high-Z solid targets. Due to the low luminosity of $\gamma$-rays, the polarized positrons obtained so far are limited to $10^4$-level per shot with a yield efficiency on the order of $10^{-5}$ $e^+/e^-$, although these may be improved in the future.

The rapid development of high-power ultra-intense laser technology [19, 20] makes it possible to generate high-energy dense positron beams via the multiphoton Breit-Wheeler (BW) process [21-23]. However, positron beams generated by previously proposed methods typically do not have spin polarization or spin resolution [24-34]. In order to attain spin-polarized positrons, several methods by use of high-intensity asymmetric laser fields colliding with multi-GeV electrons or polarized $\gamma$-photons have been proposed recently [35-38], which give only about $10^6$-level positrons per shot with an efficiency of less than $10^4$/J. Currently, it is challenging to attain such strong asymmetric laser pulses and/or polarized multi-GeV electrons/photons. In addition, these methods are very sensitive to the spatio-temporal alignment accuracy of laser pulses and electron beams and depend heavily on their parameters. It was also proposed that the electron beams can be polarized via radiative spin flips with such ultra-intense lasers [39-42]. Very recently, it was proposed to improve the generation of polarized positrons by utilizing 100PW-class laser-solid interaction configurations [43, 44], however, their experimental implementation is difficult due to the lack of such laser facilities in the near future. Moreover, these proposed approaches usually suffer from strong laser prepulse effects and large beam divergences, which become more severe at high laser intensity above $10^{22}$W/cm$^2$ with tight focusing [45]. It is a great challenge for existing methods to attain the efficiency of $10^8$/J level polarized positrons. On the other hand, recent advances in state-of-the-art accelerators [46-48] have stimulated great interest in strong-field QED studies with high-energy dense electron beams, such as QED cascades, high-energy photon emission and filamentation instabilities [49-54]. It is not yet clear so far, however, whether it is possible to efficiently generate polarized positrons driven by use of a single high-energy electron beam. Studying this problem will give insight into the beam-target interactions in a new regime and provide an effective mechanism for achieving high-energy dense polarized positron sources.

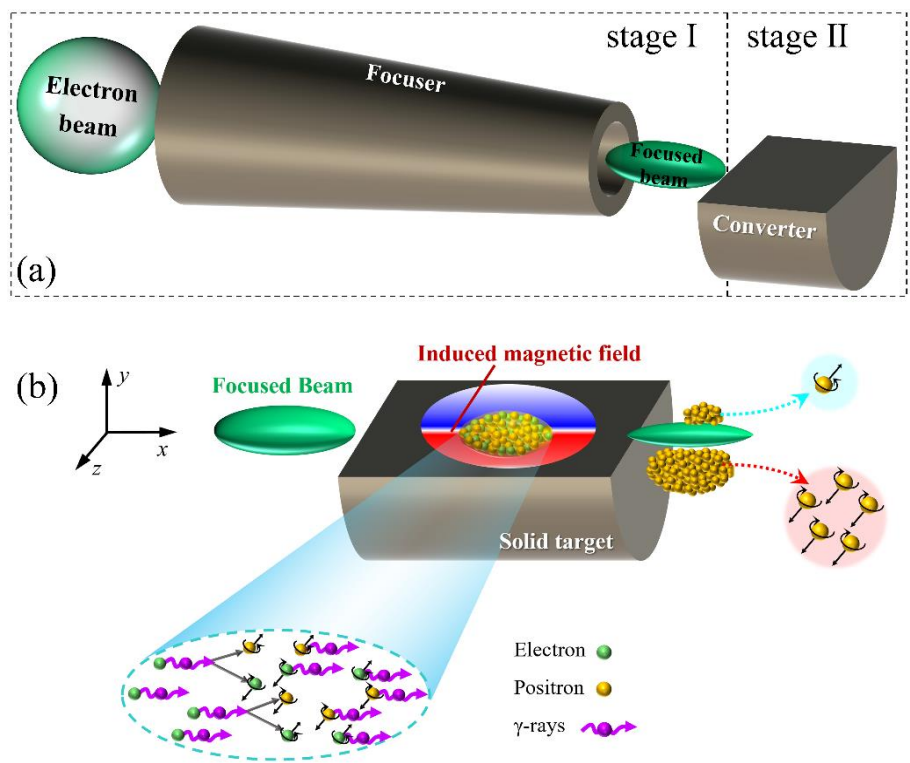

**Fig. 1.** Schematic illustration. (a) This scheme involves two stages, i.e., a relativistic electron beam is first focused by passing through a hollow cone target (the focuser) in the first stage, and subsequently the focused beam impinges the surface of a solid target (the converter) for positron generation in the second stage. (b) When such a dense beam hits the solid target surface, it excites asymmetric intense magnetic fields of megatesla magnitude at the target surface, producing a large number of energetic positrons via the multiphoton BW process. Most positrons are polarized in such intense asymmetric fields due

to the radiative spin flip effect and split into two parts in space along the y direction due to the Lorentz force.

In this Letter, a new scheme is proposed for efficient generation of polarized multi-GeV positrons by the interaction of an ultra-relativistic unpolarized electron beam with solid targets, which involves two stages, i.e., the beam focusing with a designed hollow cone target in the first stage and the subsequent interaction of the focused beam with a solid surface as the converter for polarized position generation in the second stage, as shown schematically in Fig. 1. The beam focusing to a density close to a solid target density is essential to trigger QED processes during the beam interaction with a solid target, since modern accelerators cannot yet produce beams with such a high density. It is found that when such a dense beam impinges on a solid surface, asymmetric intense magnetic fields are induced near the target surface due to the large plasma electron backflows, which triggers the multiphoton BW process to produce a large number of energetic positrons. Because the probability of spin-resolved pair production is intrinsically asymmetric and there are asymmetric field effects, most positrons are polarized via radiative spin flips. As a consequence, a multi-GeV dense positron beam with a high spin polarization can be efficiently produced.

We demonstrate the feasibility of the scheme with the polarized-QED particle-in-cell (PIC) code KLAPS [43, 45, 55], including nonlinear Compton scattering and the multiphoton BW process with pair spin and photon polarization effects, which has been fully benchmarked. These two processes are characterized by two quantum invariant parameters [56], i.e., $\chi_e = \left(\frac{e\hbar}{m_e^3 c^4}\right)\left|F_{\mu\nu}p^\nu\right|$ to determine the $\gamma$-ray photon emission and $\chi_\gamma = \left(\frac{e\hbar^2}{2m_e^3 c^4}\right)\left|F_{\mu\nu}k^\nu\right|$ to determine the electron-positron pair creation, where $e$ is the elementary charge, $m_e$ is the electron mass, $c$ is the speed of light in vacuum, $\hbar$ is the reduced Planck constant, $p^\nu(\hbar k^\nu)$ is the four-momentum of electrons or positrons (photons), and $F_{\mu\nu}$ is the field tensor. Moreover, in the simulation pair spin and photon polarization are resolved to include radiative spin flips [36, 57], and the classical spin precession is calculated according to the Thomas-Bargmann-Michel-Telegdi equation [58].

In the three-dimensional (3D) PIC simulation, we employ a window of $4\mu m(x) \times 5\mu m(y) \times 5\mu m(z)$ with grid cells of $200 \times 250 \times 250$ moving at the speed $c$ along the $x$ axis, where the macroparticles per cell for the beam electrons, plasma electrons and ions are 27, 8 and 8, respectively. As an example, we take an electron beam moving along the $x$ axis, which has about 2.8nC charge, 20GeV mean energy, 5% energy spread, and 4 mm-mrad normalized emittance. The beam has a Gaussian distribution of $\exp[-r^2/\sigma_\perp^2 - (x-vt)^2/\sigma_\parallel^2]$, where $\sigma_\perp = 2\mu m$ and $\sigma_\parallel = 1\mu m$ are used to reduce the computing resources. It is worth noting that our scheme is suitable for driving beams with different energies, such as the case with a 10GeV beam shown in the Supplemental Material [59] (including Refs. [60-67]). Comparable beam parameters will be available in some advanced accelerators in the near future [47, 48], and higher energy beams may be achieved via a combination of FACET-II with LCLS linac [49, 50]. On the other hand, although the beam already has a high density with the beam parameters given above, it is not yet high enough to trigger QED effects while interacting with a solid target. To increase the beam density further, it can be further focused by passing through a hollow cone target. For example, with a properly designed hollow cone, as described in the Supplemental Material [59], it can be self-focused to a radius of $\sim 0.4\mu m$ and its density is increased by about two orders of magnitude to close to the solid density according to our previous studies [68]. Therefore, the results shown below have been obtained based upon integrated 3D PIC simulations with the configuration given in Fig. 1(a), involving beam focusing and polarized positron generation via the interaction of the focused high-density beam with a solid target. In the following, we mainly focus on the processes of positron production and spin polarization, as illustrated in Fig. 1(b). The solid conversion target has a thickness of $L_0 = 20\mu m$ and an initial density of $n_{p0} = 1 \times 10^{29} m^{-3}$, which is attached at the cone apex. If necessary, a small distance between the hollow cone apex and the converter is also acceptable. Note that the underlying mechanisms are insensitive to the material, thus a target with other material can also be applied.

Figure 2 illustrates the mechanism and process for polarized positron generation by direct interaction of the relativistic dense electron beam with the solid target. For a relativistic electron beam propagating in vacuum, the electric term of its self-fields is almost cancelled out by its magnetic term naturally, such that the QED effects cannot be triggered. However, for the beam-target interaction in our configuration, the two fields are no longer cancelled out [compare Fig. 2(a) and Fig. 2(b)]. This forms a strong effective field $\mathbf{E}_{\text{eff}} = \mathbf{E}_\perp + \boldsymbol{\beta} \times \mathbf{B}$ on the target surface [see Fig. 2(c)], where the transverse electric field $\mathbf{E}_\perp$ and the magnetic field $\mathbf{B}$ are nearly perpendicular to the electron velocity $\boldsymbol{\beta} = \mathbf{v}/c$. The main reason is that when the dense beam impinges on the surface of a solid target, it causes a large backflow current of the plasma electrons [Fig. 2(d)]. It gives rise to a strong self-generated magnetic field on the order of megatesla, while the beam space-charge field decays rapidly within the skin depth and is thus shielded at the target surface. The magnetic field force experienced by the beam inside the target can exceed its electric field force, thus pushing the beam away from the target, and usually in such a magnetic-field-dominated (MFD) region, the QED processes occur efficiently as shown in laser-based schemes [43, 69, 70]. The induced magnetic field is related to the backflow current density $\mathbf{J}_p = -e n_p \mathbf{v}_p$ of the plasma electrons, which can be described approximately by $\nabla \times \mathbf{B} = \frac{4\pi}{c} \mathbf{J}_p$. The space charge field $E_b \propto -4\pi e n_b \sigma_\perp^2 / r$ generated by the driving beam with high density $n_b$ is large enough to expel almost all of the plasma electrons away from the target, leaving behind the massive ions there. This causes the beam electrons outside the solid target to be attracted into the target by the electrostatic Coulomb field of ions. Under the combined action of the Coulomb field and the induced magnetic field, a strong pinching force is formed [see Fig. 2(c)], making the beam confined and focused well along the target surface, as displayed in Fig. 2(e). The beam is dense enough to excite the solid surface wake and is always located in the highest field region, allowing high-efficiency intense radiation processes to occur, where the conversion efficiency into γ-ray emission can reach 65% and pair creation efficiency can reach 4%. This indicates that our scheme is in a new interaction regime, completely different from previous beam-target interactions [71, 72].

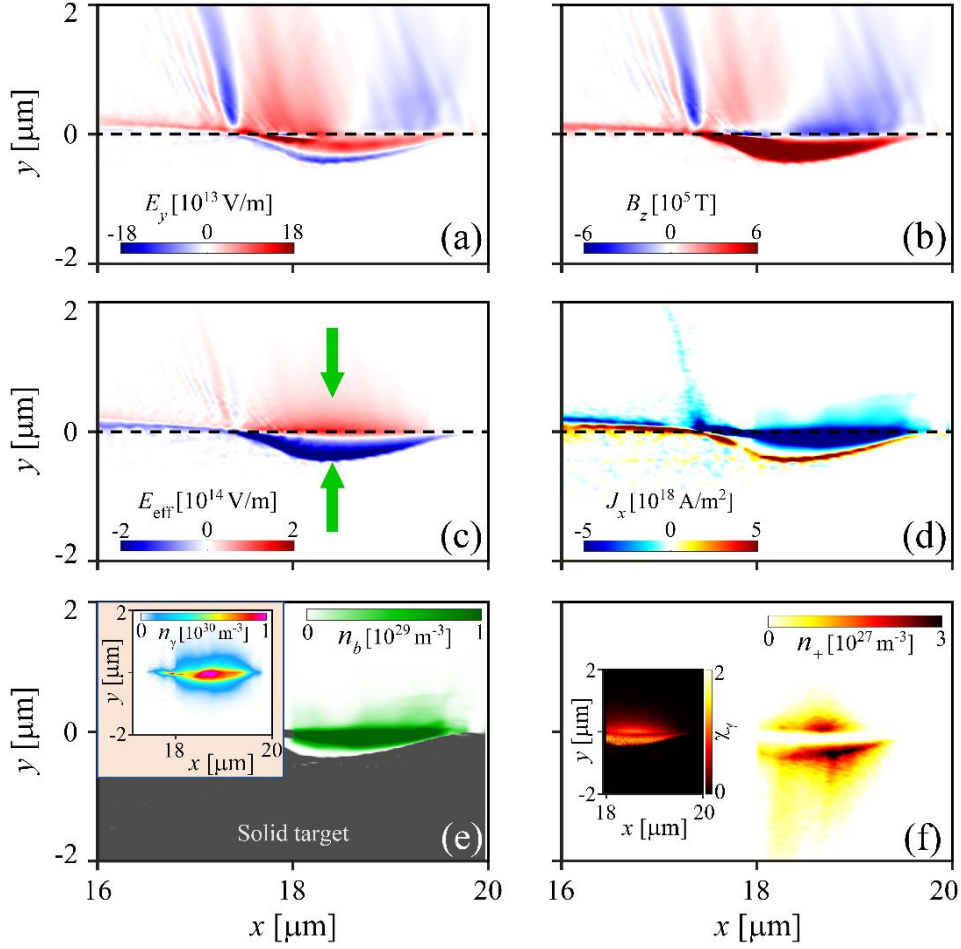

**Fig. 2.** The spatial distributions in the $(x, y)$ plane are shown at the end of the solid target: (a) the electric field $(E_y)$, (b) the magnetic field $(B_z)$, (c) the effective field $(E_{\text{eff}})$, (d) the longitudinal electric current density $(J_x)$, (e) the electron-beam density $(n_b)$, and (f) the positron density $(n_+)$. Here the black dashed line represents the initial target surface located at $y = 0$. In (c), the green arrows indicate the direction of the pinching force on the electron beam. The inset in (e) displays the $\gamma$-photon density distribution $(n_\gamma)$, and the inset in (f) exhibits the QED parameter distribution $(\chi_\gamma)$.

In the high-strength MFD region with $|\mathbf{E}_{\text{eff}}| > 2 \times 10^{14} \text{V/m}$, strong-field QED effects can be efficiently triggered from direct electron beam-target interaction without the use of high-intensity laser pulses. For 20GeV electron beam, the quantum parameter of photon emission can be as high as $\chi_e = \gamma_0 |\mathbf{E}_{\text{eff}}|/E_S \sim 10$, where $\gamma_0$ is the relativistic Lorentz factor, and $E_S \approx 1.3 \times 10^{18} \text{V/m}$ is the critical field of QED [73]. Hence, large amounts of high-energy $\gamma$-rays are emitted at an unprecedented high density, as seen in the inset of Fig. 2(e). These high-energy photons then decay into electron-positron pairs through the multiphoton BW process. The magnetic field generated inside the target is much greater than the field generated outside the target, which leads to the difference in pair yields dominated by the characteristic QED parameter $\chi_\gamma = (\hbar\omega/2m_ec^2E_S)|\mathbf{E}_\perp + \hat{\mathbf{k}} \times \mathbf{B}| \propto |\mathbf{B}|$ in these two regions, where $\hbar\omega$ is the emitted photon energy with the unit wave vector $\hat{\mathbf{k}}$ along the parent lepton (electron or positron) velocity. Accordingly, the number of positrons born in the $y < 0$ region is much higher than that born in the $y > 0$ region, which is demonstrated by our PIC simulations as shown in Fig. 2(f). Furthermore, unlike electrons guided by the effective surface field, this field here tends to separate positively charged particles (say, positrons here), such that the positrons are divided into two parts, one in the $y < 0$ region and the other in the $y > 0$ region. In such asymmetric fields, the created positrons can break the polarization symmetry, and they split into the two regions with opposite polarizations because of the angle-dependent polarization, making highly polarized positrons possible, as explained in detail below.

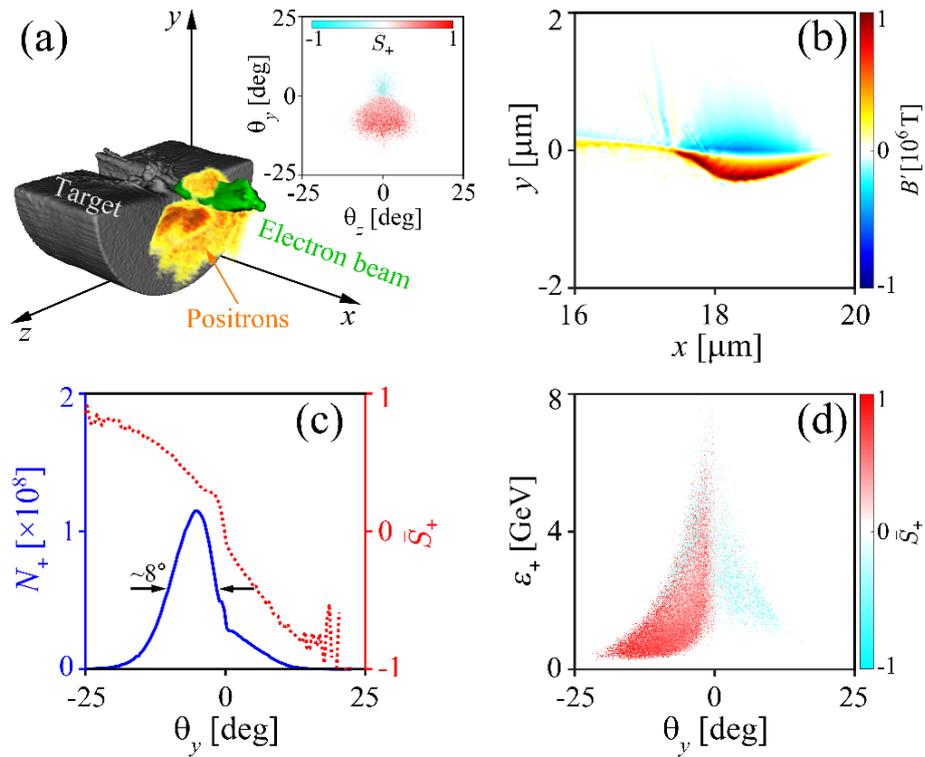

**Fig. 3.** (a) 3D view of polarized positrons produced by the beam-target interaction, where the inset shows the angular distribution of the positron polarization $\bar{S}_+$. (b) The spatial distribution of the magnetic field $\mathbf{B}'$ in the $(x, y)$ plane. (c) The positron yield $N_+$ and polarization $\bar{S}_+$ as a function of the angle $\theta_y$. (d) The distribution of the positron polarization versus the energy $\varepsilon_+$ and angle $\theta_y$.

In order to elucidate the underlying physics of positron polarization formation, we investigate the effect of asymmetric magnetic fields on the spin dynamics in Fig. 3. It is shown that in addition to determining the generation of positrons, the induced asymmetric fields also determine their polarization and deflection. When high-energy photons are converted into electron-positron pairs via the multiphoton BW process, the spin vector $\boldsymbol{S}_+$ of the newborn positron can be set to $\boldsymbol{S}_+^*/|\boldsymbol{S}_+^*|$ [74], where $\boldsymbol{S}_+^* = -\xi_1' \frac{\varepsilon_\gamma}{\varepsilon_-} K_{1/3}(y_2)\boldsymbol{e}_1' + \xi_2' \left[ \frac{\varepsilon_\gamma}{\varepsilon_+} \text{Int}K_{1/3}(y_2) + \left( \frac{\varepsilon_+}{\varepsilon_-} - \frac{\varepsilon_-}{\varepsilon_+} \right) K_{2/3}(y_2) \right] \boldsymbol{e}_\nu' - \left( \frac{\varepsilon_\gamma}{\varepsilon_+} - \xi_3' \frac{\varepsilon_\gamma}{\varepsilon_-} \right) K_{1/3}(y_2)\boldsymbol{e}_2'$, $y_2 = 2\varepsilon_\gamma^2/3\chi_\gamma\varepsilon_+\varepsilon_-$, $\text{Int}K_{1/3}(y) \equiv \int_y^\infty K_{1/3}(x)dx$, $K_n(y)$ is the $n$-order modified Bessel function of the second kind, $\varepsilon_\gamma$ is the photon energy, and $\varepsilon_+$ ($\varepsilon_-$) is the positron (electron) energy. The photon polarization can be characterized by Stokes parameters $\boldsymbol{\xi} = (\xi_1, \xi_2, \xi_3)$, defined with respect to the basis vector $(\boldsymbol{e}_1, \boldsymbol{e}_2, \boldsymbol{e}_\nu)$, where $\boldsymbol{e}_\nu$ is the unit vector along the lepton velocity, $\boldsymbol{e}_1$ is the unit vector along the transverse acceleration, and $\boldsymbol{e}_2 = \boldsymbol{e}_\nu \times \boldsymbol{e}_1$. Based on the matrix rotation [75], the Stokes parameters $\boldsymbol{\xi} = (\xi_1, \xi_2, \xi_3)$ can be transformed from the photon emission frame $(\boldsymbol{e}_1, \boldsymbol{e}_2, \boldsymbol{e}_\nu)$ to the pair creation frame $(\boldsymbol{e}_1', \boldsymbol{e}_2', \boldsymbol{e}_\nu')$ to give $\boldsymbol{\xi}' = (\xi_1', \xi_2', \xi_3')$, where $\boldsymbol{e}_1'$ is the unit vector along $\boldsymbol{E} + \boldsymbol{e}_\nu \times \boldsymbol{B} - \boldsymbol{e}_\nu(\boldsymbol{e}_\nu \cdot \boldsymbol{E})$, $\boldsymbol{e}_2' = \boldsymbol{e}_\nu \times \boldsymbol{e}_1'$, $\boldsymbol{e}_\nu = \boldsymbol{e}_\nu'$, and $\theta$ is the angle between $\boldsymbol{e}_1$ and $\boldsymbol{e}_1'$. The resulting positrons are more likely to have spins parallel to the magnetic field direction $\boldsymbol{\zeta} \equiv \gamma_0 \boldsymbol{B}'/|\gamma_0 \boldsymbol{B}'|$ in their rest frames, where $\boldsymbol{B}' \approx \boldsymbol{B} - \boldsymbol{\beta} \times \boldsymbol{E} - \boldsymbol{\beta}(\boldsymbol{\beta} \cdot \boldsymbol{B})$ in the relativistic limit of $\gamma_0 \gg 1$, and the unit vector $\boldsymbol{\beta}$ along the positron velocity is approximately perpendicular to the magnetic field. Considering that $\boldsymbol{\beta}$ mainly points to the $x$-$y$ plane, one can further obtain $\boldsymbol{\zeta} \approx (0, 0, B_z/|B_z|)$ in the MFD regime. Therefore, the average polarization $\bar{S}_+$ of the newborn positrons is basically in the same direction as $B_z$, along the z direction, as observed in our PIC simulation results [see Figs. 3(a) and 3(b)].

In such asymmetric fields, the positrons are deflected by the Lorentz force $\boldsymbol{f}_+ = q\mathbf{E}_{\text{eff}}$ towards the region for $y < 0$ with $f_+ < 0$ or towards the region for $y > 0$ with $f_+ > 0$, where $q = +e$, and the effective field $\mathbf{E}_{\text{eff}}$ is presented in Fig. 2(c). Under the combined action of the Lorentz force and radiative spin flip, positrons produced in the region of $y < 0$ with $B_z > 0$ are mostly polarized with $\bar{S}_+ > 0$ and deflected toward $\theta_y < 0$; While positrons produced in the region of $y > 0$ with $B_z < 0$ are mostly polarized with $\bar{S}_+ < 0$ and deflected toward $\theta_y > 0$. As a result, the transversely polarized positrons with spin polarization along $\pm z$ directions are deflected along $\mp y$ directions. Because of the stronger magnetic field excited inside the target (that is the MFD region), more than 85% of the total positrons are generated in the $\theta_y < 0$ region, with the average polarization reaching approximately 42%, as illustrated in Fig. 3(c). When selecting positrons in the angle range of $\theta_y <$ $-15°$, the spin polarization will reach 80%, accounting for about 3% of the total number of positrons.

On the other hand, the findings indicate that the polarization of positrons also depends on their energies, where the positrons have higher polarization in the lower energy range, as shown in Fig. 3(d). We elucidate the physics involved as follows. Physically, when emitting photons with low energy $\varepsilon_\gamma \ll \varepsilon_e$, the parent lepton can keep its polarization nearly unchanged (that is $S_f \to S_i$), and the average polarization of emitted photons always has a positive value of $\bar{S}_\gamma \to 0.5$, as detailed in Supplemental Material [59]. Here $S_i$ and $S_f$ are the spin states of the lepton before and after photon emission, and $\varepsilon_e$ is the emitting lepton energy. While for emitting high-energy photons with $\varepsilon_\gamma/\varepsilon_e \to 1$, the photon polarization is highly dependent on lepton polarization, that is $\bar{S}_\gamma \to S_i$, where the spin vector of leptons after photon emission tends to be parallel (antiparallel) to the magnetic field direction $\boldsymbol{\zeta}$ for positrons (electrons). Therefore, when strong radiation effects occur, high-energy positrons tend to flip their spin vectors along the magnetic field direction and lose a lot of energy to emit high-energy photons, so they become highly polarized low-energy positrons; Otherwise, their spin polarization changes little after creation and is primarily determined by the polarization of their parent photons. Since γ-ray photons emitted in the high energy region have lower spin polarization, when they are converted into pairs, positrons born in the high-energy region are less polarized than those born in the low-energy region; see Supplemental Material [59] for more details. These are the main reasons why positron polarization is relatively small in the high energy range and relatively large in the low energy range.

To demonstrate the robustness of the proposed scheme, we investigate the effects of the target thickness and driving electron beam charge parameters on positron polarization $\bar{S}_+$ and yield $N_+$. We first discuss the effect of the target thickness on the positron production and polarization. It is shown that the interaction of the driving beam with a solid target of appropriate thickness is conducive to exciting stronger magnetic fields, leading to higher average polarization of positrons. For example, when using a solid target of $L_0 = 10\mu m$, the positron polarization $\bar{S}_+$ reaches approximately 45%, as seen in Fig. 4(a). With a thinner target (for example $L_0 < 5\mu m$), the beam would not have enough time to build up a strong magnetic field, causing the decrease in both positron yield and polarizaiton. However, if an thicker target is adopted (e.g., $L_0 > 20\mu m$), the electron beam loses a significant amount of energy and thus attenuates, causing the induced magnetic fields to weaken later in the interaction. In addition, the deflection effect of the Lorentz force prevents the newborn positrons from staying in the high-intensity region of the induced magnetic fields for a long time, so the positrons will not be well polarized. As a result, more positrons are produced, with a number increased to $5.1 \times 10^9$ (about 0.8nC charge) in the case of $L_0 = 40\mu m$, but their polarization decreases to 36%. Finally, it gives an unprecedented high efficiency of up to $10^8$ positrons/J, with the yield of about 0.3 $e^+/e^-$, which cannot be achieved by other methods in the prior art.

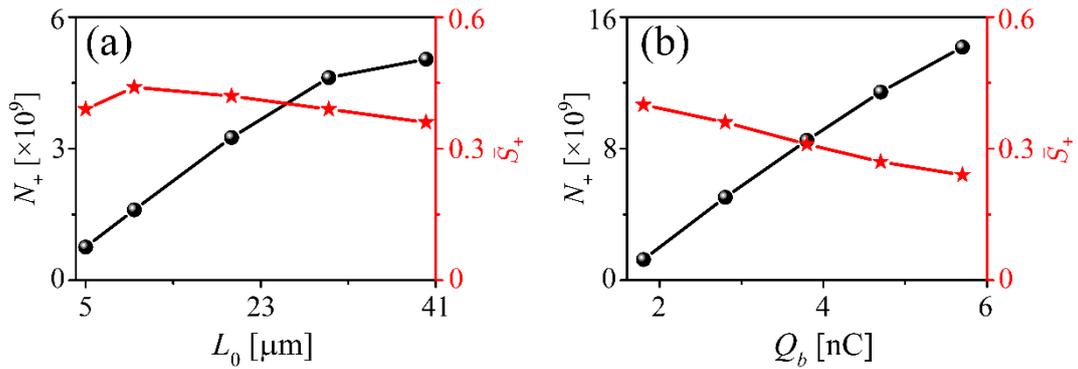

**Fig. 4.** The positron yield $N_+$ and polarization $\bar{S}_+$ as functions of (a) the target thickness $L_0$ and (b) driving electron beam charge $Q_b$.

Figure 4(b) presents the dependence of positron polarization $\bar{S}_+$ and yield $N_+$ on the driving electron beam charge $Q_b$. It indicates that the yield of positrons created increases with the growth of beam charge, while their polarization tends to decrease. In fact, a higher charge electron beam can drive larger plasma static fields and trigger stronger QED effects, producing a large number of polarized positrons, but they may mix with each other, leading to a decrease in the average spin polarization of the total positrons. Furthermore, since positrons created in such large fields undergo stronger Lorentz forces, they will quickly be deflected and leave the high-intensity region of the induced magnetic fields. In this case, although the number of newly created positrons increases significantly, there is not enough interaction time for the positrons to gain high spin polarization. For example, when the driving electron beam has a charge of $Q_b = 5.7\text{nC}$, the total charge of the obtained positrons is as high as 2.3nC, and the average spin polarization is approximately 24%; When $Q_b = 1.8\text{nC}$, the obtained positron charge is reduced to 0.2nC, while the average polarization is considerably increased to about 40%. Overall, the polarization and charge of high-energy positrons can be tuned by simply changing the target and electron beam parameters to meet different requirements.

For the experimental realization of our scheme, in order to achieve a high repetition rate in experiments, a movable target tape can be employed, where the targets are positioned at the focus of the driving beam through a tape target delivery system. This method has been achieved with current technology and used in many studies of laser plasma interactions [76, 77]. In addition, the involved quasi-static electric and magnetic fields are induced by the background target electrons in the timescale $\tau_e = 1/4\pi\sigma$ on the order of $10^{-18}s$ for common conductors [78], where $\sigma$ is the electrical conductivity. Since $\tau_e \ll \tau_b$ (where $\tau_b \sim 10\text{fs}$ is the beam duration), intense quasi-static magnetic fields are generated fast enough to focus the beam as found in our simulation. In passing, one notes that miniature magnetic devices or plasma lens [79, 80] may be used to control the transport and focusing of relativistic positrons for collection and applications.

To summarize, we have proposed a simple approach to efficiently produce dense polarized positrons through electron-beam-solid interactions without the use of high-intensity lasers. It is shown that multi-GeV polarized positrons with nC-scale charge and high spin polarization above 40% can be produced by impinging a relativistic dense electron beam on the solid target surface. Since the induced magnetic fields experienced by the positrons are naturally asymmetric, the spin polarization mechanism is robust, making it feasible to generate polarized positrons in realistic beam-solid interactions. By increasing the driving beam charge and/or energy, more positrons can be produced. Such polarized positron sources may open the door to many research frontiers, such as exploring possible new physics beyond the standard model and spin-polarization related particle physics [6, 81], high-field QED physics [2, 3], and laboratory astrophysics [8-10].

**Acknowledgements**

This work was supported by the National Natural Science Foundation of China (12205186, 12135009, 11991074, 11975154, and 11775302), and the Research Funds of Renmin University of China (20XNLG01).